\documentstyle[prl,aps,epsfig,twocolumn,amsmath]{revtex}        
\newcommand{\upd}{{\mathrm d}}
\renewcommand{\epsilon}{\varepsilon}

\begin{document}

\draft
\twocolumn[\hsize\textwidth\columnwidth\hsize\csname @twocolumnfalse\endcsname
\title{ \bf Phase separation in a chaotic flow}
\author{ Ludovic Berthier$^{\star,\star\star}$,
Jean-Louis Barrat$^\star$
and Jorge Kurchan$^{\star\star\star}$}  

\address{$^\star$D\'epartement de Physique des Mat\'eriaux, 
Universit\'e C.
Bernard and CNRS, F-69622 Villeurbanne, France} 

\address{$^{\star\star}$Laboratoire de Physique, ENS-Lyon and CNRS,
 F-69007 Lyon, France }

\address{$^{\star\star\star}$PMMH, \'Ecole Sup\'erieure
de Physique et Chimie Industrielles,
F-75005 Paris, France}  

\date{\today}

\maketitle

\begin{abstract}
The phase separation between two immiscible liquids advected
by a bidimensional  velocity field is investigated numerically 
by solving the corresponding Cahn-Hilliard equation. 
We study how the  spinodal decomposition process  depends 
on the presence --- or absence --- of Lagrangian chaos.
A fully chaotic flow, in particular, limits the growth
of domains and 
for unequal volume fractions of the liquids, a 
characteristic exponential distribution of droplet sizes is obtained. 
The limiting domain size results from a balance
between chaotic mixing and spinodal decomposition, measured in terms 
of Lyapunov exponent and diffusivity constant, respectively.
\end{abstract}  

\pacs{PACS numbers: 05.70.Ln, 47.55.Kf, 64.75.+g  \hspace*{4.3cm} 
LPENSL-TH-12/2000}

% 05 Statistical physics and thermodynamics
% 05.70.Ln Nonequilibrium thermodynamics, irreversible processes  

% 47. Fluid Dynamics
% 47.55.Kf Multiphase and particle-laden flows

% 64 Equations of state, phase equilibria, and phase transitions
% 64.75.+g Solubility, segregation, and mixing; phase separation    

\vskip2pc]

\narrowtext

 A system  of two immiscible fluids at rest 
will gradually phase-separate,
forming  domains whose size grows  algebraically with time.
Everyday experience, however, shows  that by continuously
 stirring or shaking the fluids
the domains or droplets of the phases (say, oil
and vinegar) break and coalesce, leading to a dynamic
stationary state with domains of finite size.

A first approach consists in  modelling  this situation by subjecting
 the binary
fluid to a homogeneous shear velocity field~\cite{onuki}.
However, experiments~\cite{bands}, numerical simulations~\cite{cogola},
 and more recently
analytical approaches~\cite{Cavagna} show that in such a situation  
infinitely long
 domains aligned with the flow are formed. The effect of the
velocity field is to 
 counter  the Rayleigh instability, stabilising {\em lamellar} and
 (in certain cases)  even cylindrical domains~\cite{amalie,foot}. 
Domain breakup
in those situations takes place only at large Reynolds numbers,
 and is generally attributed
to inertial effects~\cite{onuki,Julia}. Studying these
inertial effects  numerically is difficult, as a realistic description of
the feedback of domain shape on the flow is required~\cite{Julia}. 
The corresponding
simulations are therefore limited by finite size effects.

In  this paper we investigate a different mechanism by which domains of finite
size can be stabilized in a demixing system.  
In particular,  we show 
that a saturation of the average 
length scale takes place  even in the absence of inertial effects 
if the flow has {\em Lagrangian chaos}, ({\it i.e.} if the trajectories of
nearby starting  points diverge exponentially with time). 
This is interesting for two reasons: 
First, with an appropriate time-dependence of the 
velocity field one can still have Lagrangian chaos in a 
situation of high viscosity in which inertial effects are 
negligible --- this is how one mixes pastes. 
Secondly, it is possible in that case
to decouple the hydrodynamic problem from the phase separation. 
This problem of a {\em passive, phase separating scalar field   } 
is of course much simpler, so that simulations using large systems are 
possible. Our approach therefore extends 
earlier extensive studies of 
{\em passive scalar} advection by  periodically driven 
chaotic flows  \cite{ottino}. 

Our study is also related in spirit to earlier
studies of advection by `synthetic' velocity fields tuned
to model turbulent flows \cite{siggia,babiano}. Phase separation
was studied in this context in Ref. \cite{esp}. An essential difference,
compared to our work, is that in such turbulent  flows the separation between
nearby  tracer particles appears to increase algebraically, rather than
exponentially, with time. 

\begin{figure}
\begin{center}
\begin{tabular}{cc}
\psfig{file=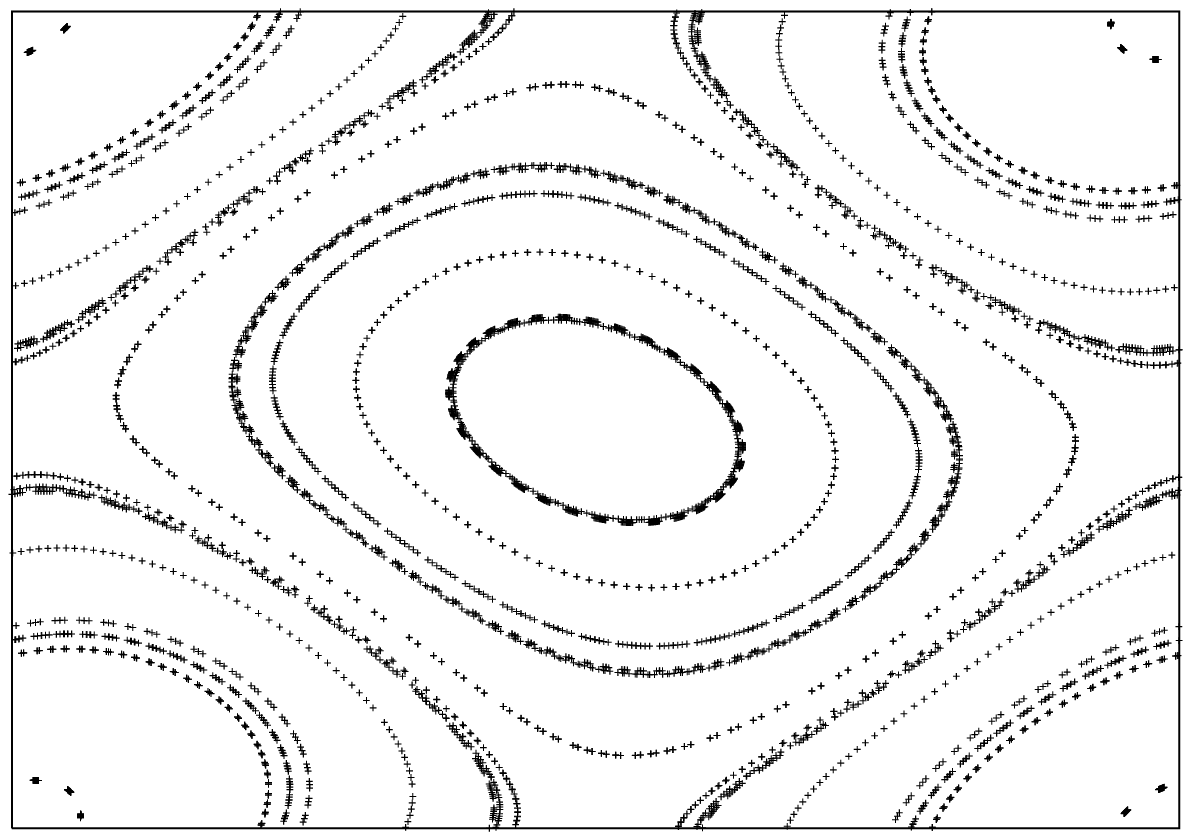,width=3.5cm,height=3.5cm} &
\psfig{file=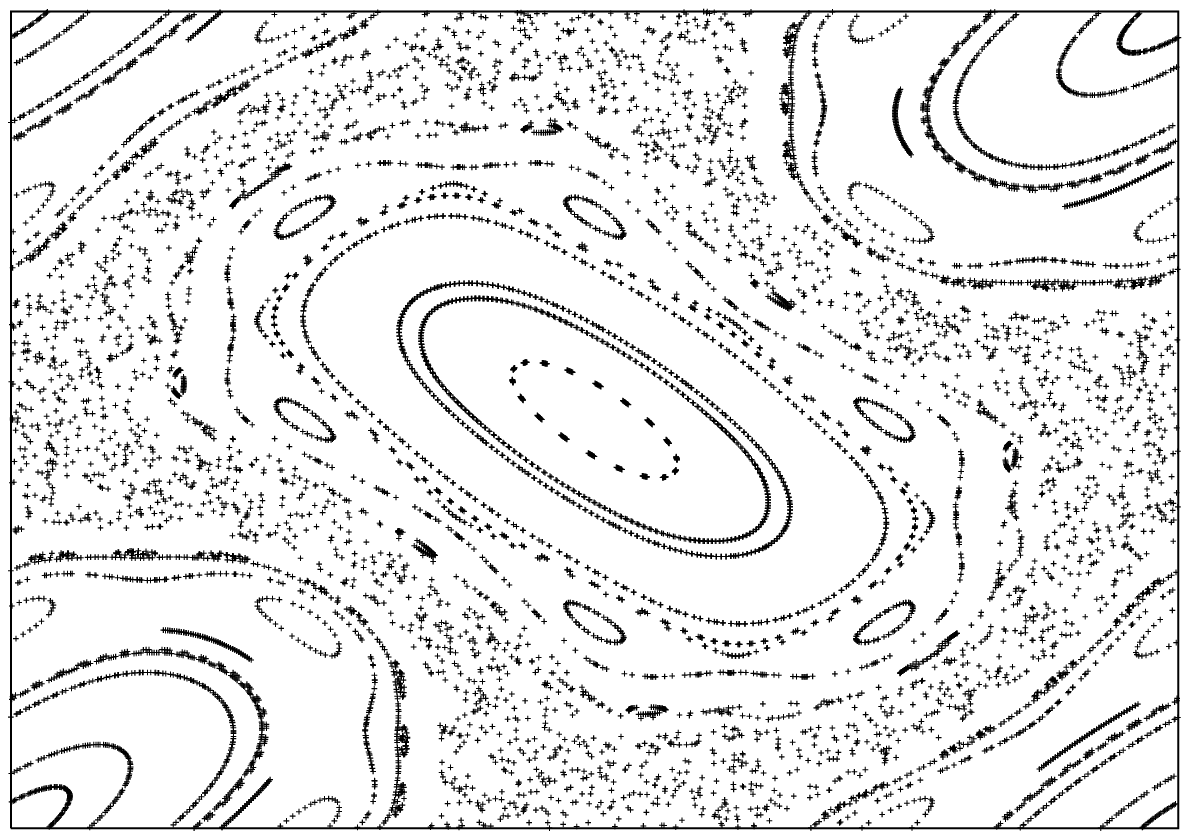,width=3.5cm,height=3.5cm} \\
\psfig{file=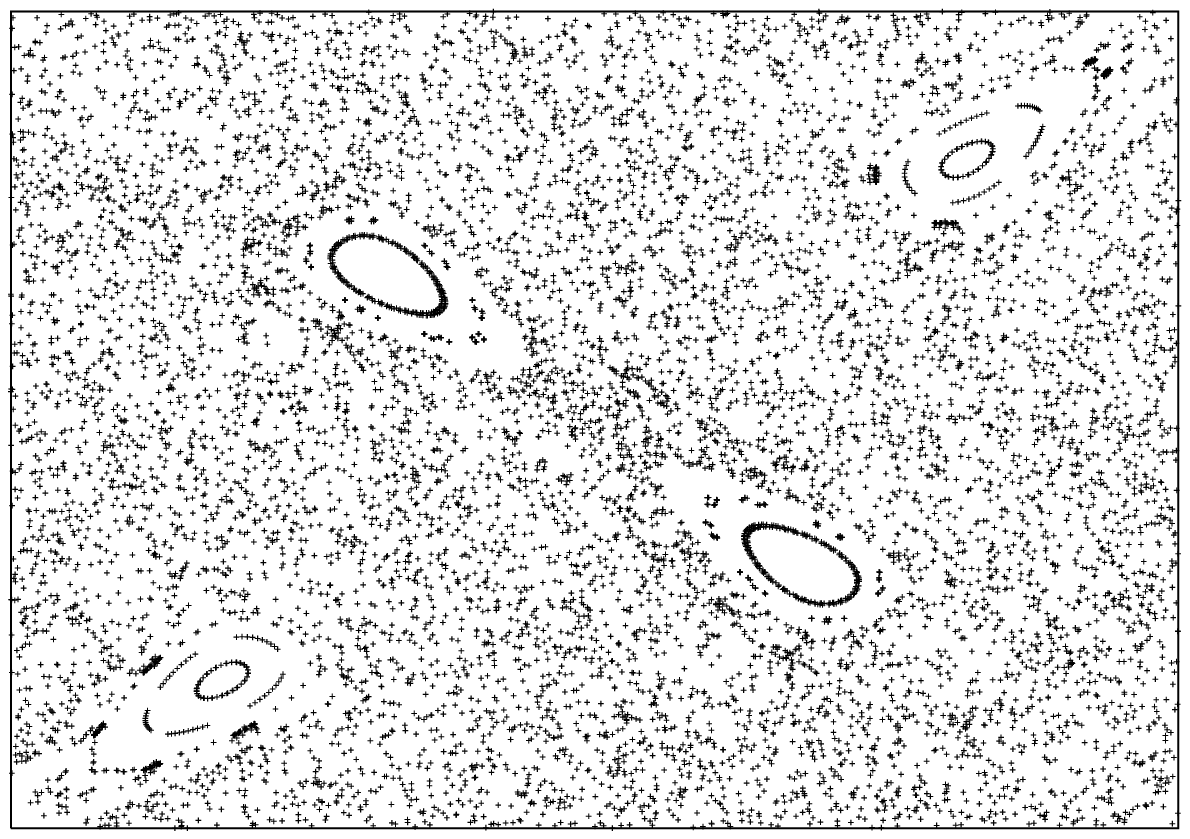,width=3.5cm,height=3.5cm} &
\psfig{file=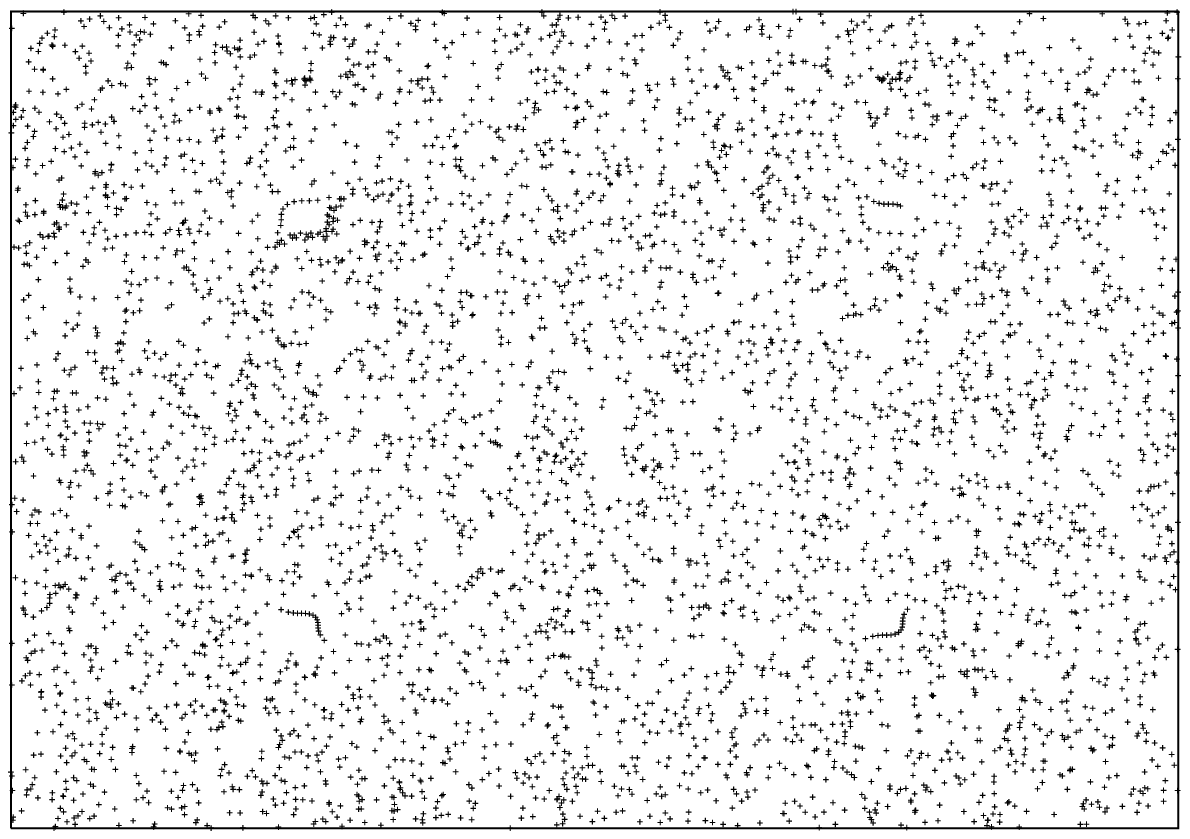,width=3.5cm,height=3.5cm}
\end{tabular}
\caption{The maps obtained from snapshots at intervals $T$ of the
  lines of current with the dynamics (\ref{dos}), starting
  from various initial conditions.
Figures for $\alpha=0.1$ (top left), 0.25, 0.4 and 1.0 (bottom right).}
\label{field}
\end{center}
\end{figure}

We consider a two dimensional
flow
 that can be tuned to be regular, mixed, 
 or fully chaotic. 
Specifically, the incompressible 
velocity field $\boldsymbol{v}(x,y,t)$ is a modified version
of the so-called  time-dependent  Harper map~\cite{Zas}
(related to the
`partitioned-pipe mixer', a special case of `eggbeater flow'~\cite{ottino}).
The dynamics takes place on a square of side $L$, with periodic
boundary conditions. The velocity field is an alternating  sequence 
of shears in
the
$x$ and in the $y$ direction with a time period $T$,
\begin{equation}
\begin{aligned}
v_x & = -\frac{2\pi \alpha L}{T} \sin \left(\frac{2\pi y}{L}\right); \;
 v_y = 0  ; &  \quad
& n < \frac{t}{T} < n+\frac{1}{2}  \\   
v_y & =
\frac{2 \pi \alpha L}{T} \sin \left( \frac{2 \pi x}{L} \right); \;
 v_x  = 0; &  
    n + & \frac{1}{2} < \frac{t}{T} < n+1.
\label{dos}
\end{aligned}
\end{equation}
The parameter $\alpha$ controls the chaoticity of the trajectories. If
 $\alpha$ is small, the two semicycles are composed into the smooth, laminar
velocity field:
$
v_x= -\frac{\pi \alpha L}{T} \sin \left(\frac{2\pi y}{L}\right)
$; $v_y =
\frac{\pi \alpha L}{T} \sin \left(\frac{2\pi x}{L}\right)$.
For larger values of $\alpha$ the trajectories stretch and fold,
and the flow becomes chaotic.
In order to visualise this, it is convenient to follow the
position of a point at the end of each cycle. This `kicked Harper' map
is shown in Fig.~\ref{field} for several values of  $\alpha$.
For $\alpha \sim 0.2$ the flow is a mixture of laminar and chaotic
regions, and becomes fully chaotic around $\alpha \sim 0.4$.
In the chaotic situation, it is convenient to characterise the flow
by the Lyapunov exponent $\lambda$, defined by the fact
that nearby starting points separate as $\sim e^{\lambda t}$.
We have computed $\lambda$  as in Ref.~\cite{babiano}
and  found that the relation $\lambda \simeq 1.96 \ln (3.35 \alpha)/T$ is
a good approximation throughout the chaotic regime, $\alpha \gtrsim 0.4$.

The spinodal decomposition of the two-component fluid is described
 by the  Cahn-Hilliard equation
\begin{equation}
 \frac{\partial \phi(\boldsymbol{r},t)}{\partial t} +
 \boldsymbol{v}(\boldsymbol{r},t)
 \cdot
\boldsymbol{\nabla} \phi(\boldsymbol{r},t)  =  \Gamma \nabla^2 \left( \frac
 {\delta F[\phi]}{\delta \phi
(\boldsymbol{r},t)} \right) .
\label{cahn}
\end{equation}
Here $\phi$ is a dimensionless concentration field, the concentrations of the
species are  $[1 \pm \phi]/2$.
 We work at $T=0$,  since temperature is irrelevant 
in this process~\cite{alan1}.
The free energy functional is of the Ginzburg-Landau form and reads
\begin{equation}
F [\phi] = \int \upd^d \boldsymbol{x} \left[ \frac{\xi^2}{2} 
(\boldsymbol{\nabla} \phi)^2 - \frac{1}{2} 
\phi^2 + \frac{1}{4} \phi^4 \right].
\end{equation}
Here,  $\xi$ is the equilibrium correlation length 
controlling  the width of the interfaces, and
$d$ is the number of spatial dimensions.
We consider two topologically  different situations
 ({\em i})   $\langle \phi \rangle \neq 0$ : a species
is less abundant than the other and forms disconnected droplets,
and  ({\em ii}) 
  $\langle \phi \rangle = 0$:  the two phases are in equal quantity
and form a bicontinuous structure. 
Situation{\em (i)} has been studied experimentally~\cite{muzio}.

In a chaotic flow, the passive scalar mixes rapidly, whereas
in the case of phase separation this
tendency is opposed by surface tension.
The competition between these two effects
can be quantified through two adimensional parameters,
$D \equiv \Gamma T/ \xi^2$ 
(the adimensional transport coefficient of the Cahn-Hilliard equation) 
and the chaoticity parameter $\alpha$,
or alternatively the adimensional Lyapounov exponent $\lambda T$.
A large $D$ means that appreciable diffusive transport will take
place during each laminar half cycle. A large $\lambda T$, on the other hand,
means that the mixing process is efficient within a few cycles. Note that
$\lambda$ can also be interpreted as an average elongation or shear
rate experienced by the fluid particles.

Equation~(\ref{cahn}) is integrated numerically with the velocity field
(\ref{dos}), using the implicit spectral method
developed and discussed in Ref.~\cite{BBK}. The results are
presented with  time and
length units chosen as the cycle period $T$ and the
interfacial thickness $\xi$, respectively.  
The system size, lattice parameter
and time step  are $L=512\xi$, $\Delta x = \xi$, and
$\Delta t = 5 \cdot 10^{-4}T$ respectively.

\begin{figure}
\begin{center}
\begin{tabular}{cc}
\psfig{file=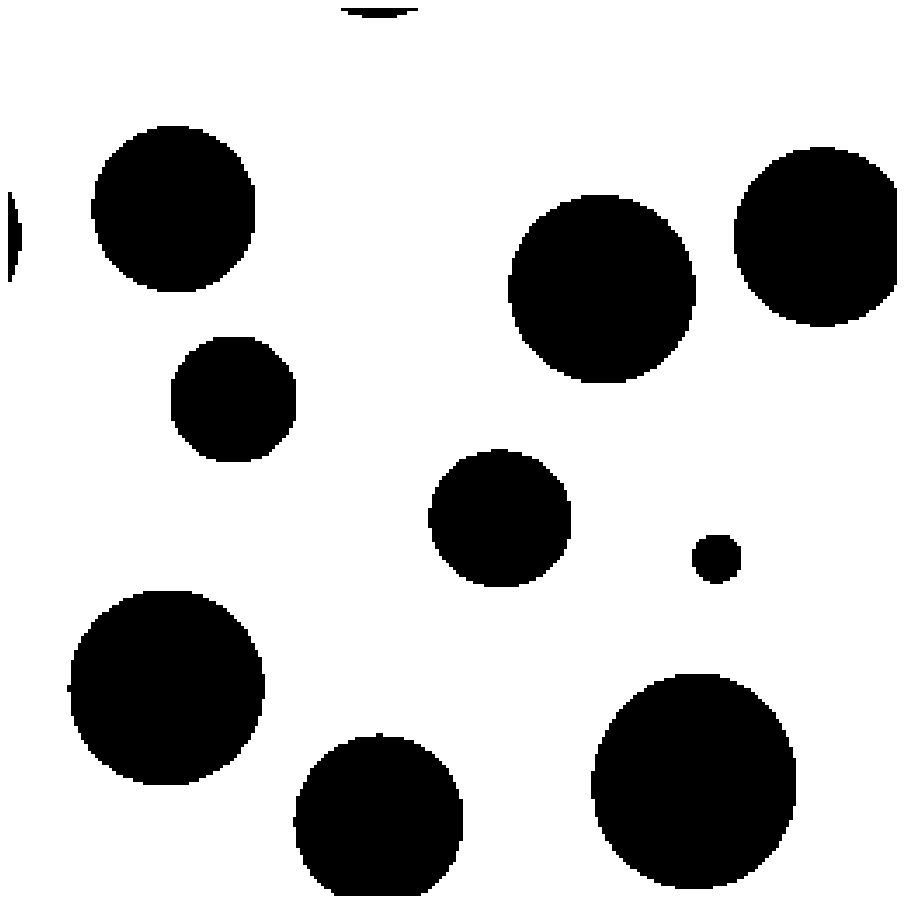,width=3.5cm,height=3.5cm} &
\psfig{file=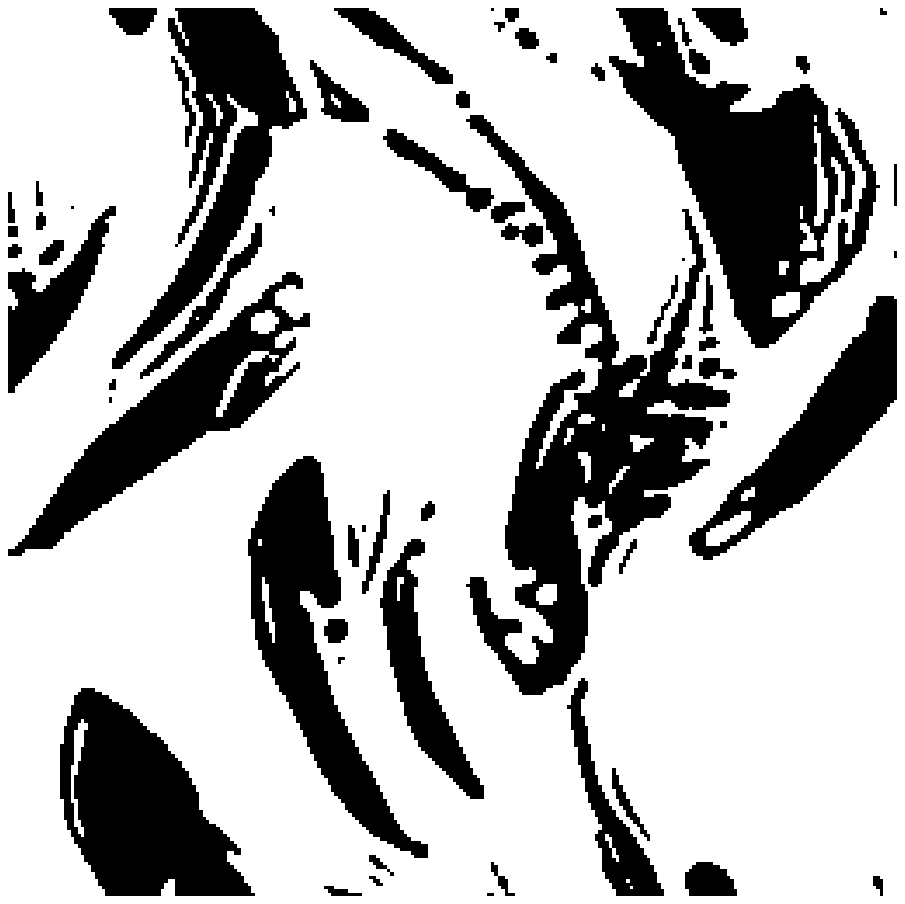,width=3.5cm,height=3.5cm} \\
\psfig{file=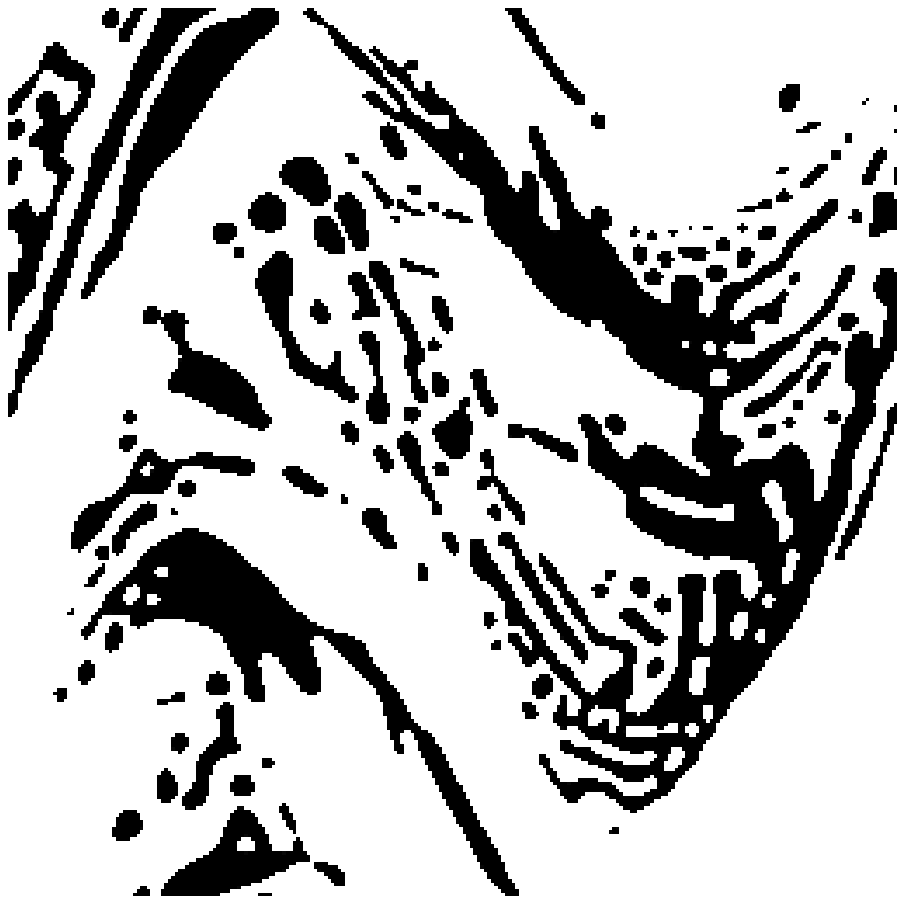,width=3.5cm,height=3.5cm} &
\psfig{file=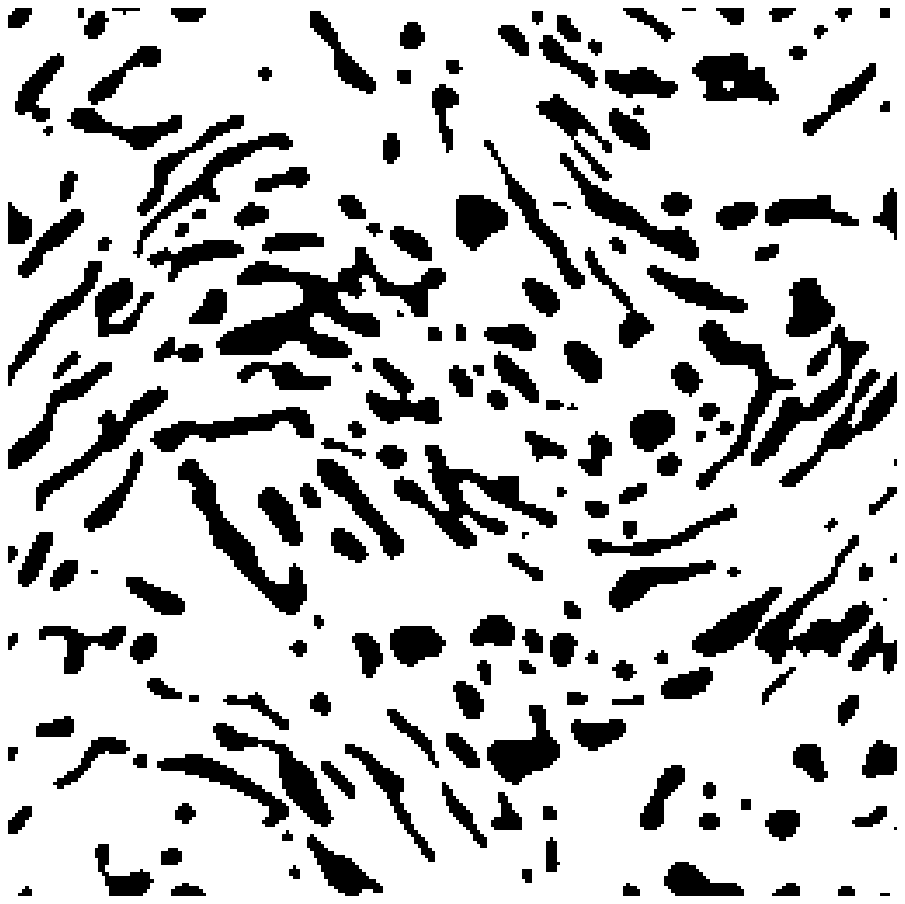,width=3.5cm,height=3.5cm}
\end{tabular}
\caption{Evolution of an assembly of large droplets in the chaotic
flow for times $t=0$, 0.65, 1.50 and 11.8.}
\label{secasse}
\end{center}
\end{figure}

\begin{figure}
\begin{center}
\begin{tabular}{cc}
\psfig{file=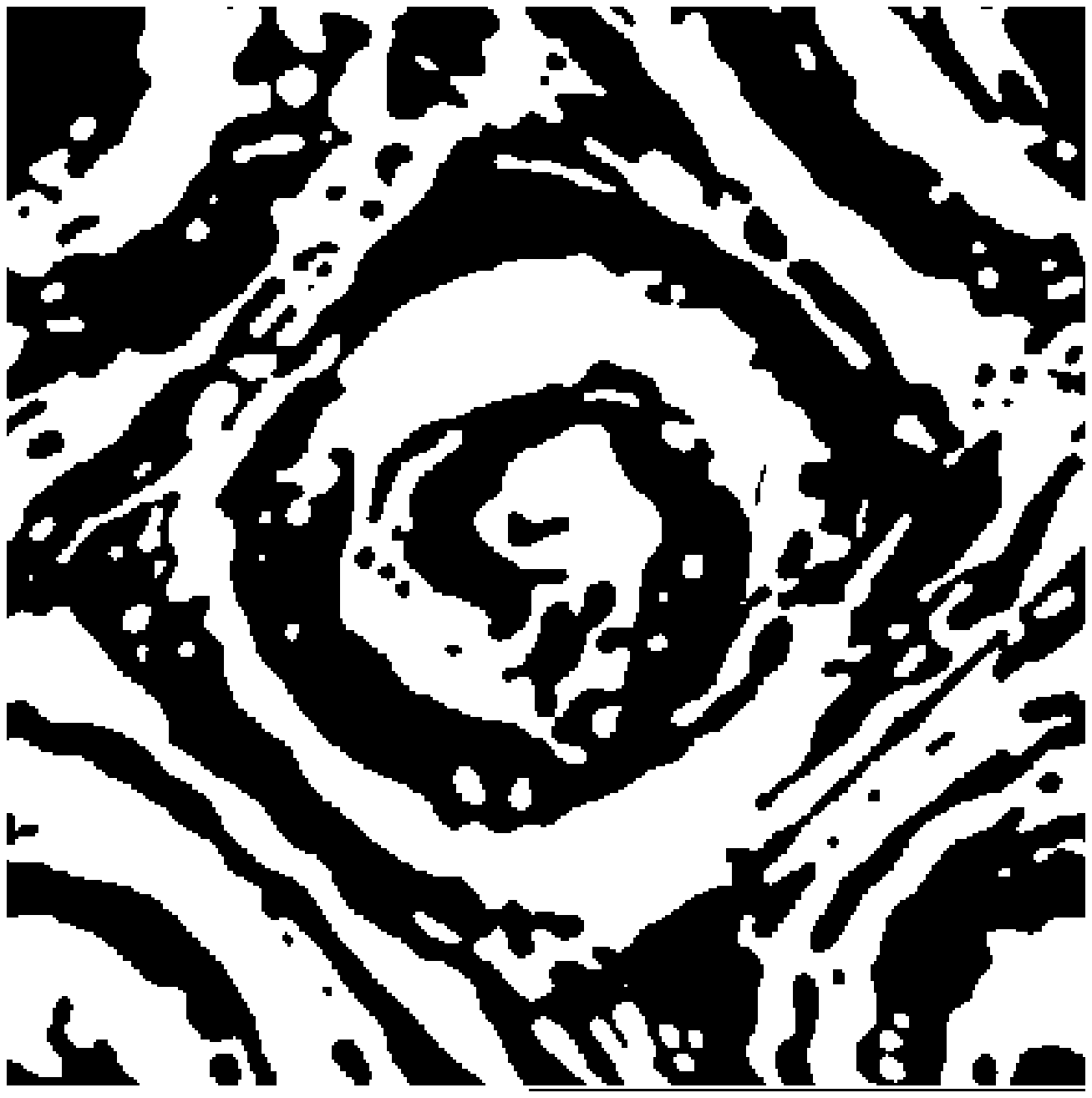,width=3.3cm,height=3.3cm} &
\psfig{file=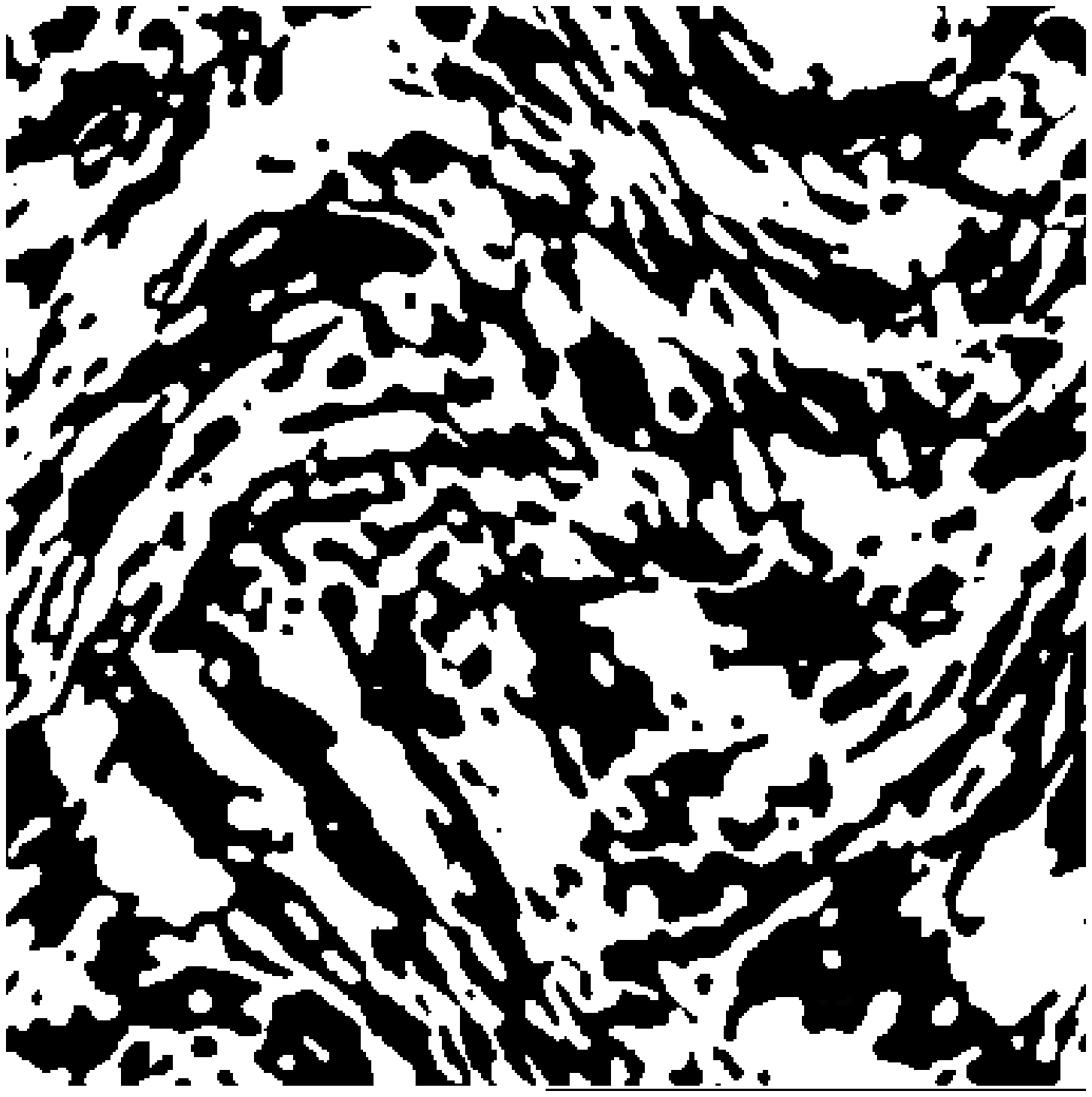,width=3.3cm,height=3.3cm} \\
\psfig{file=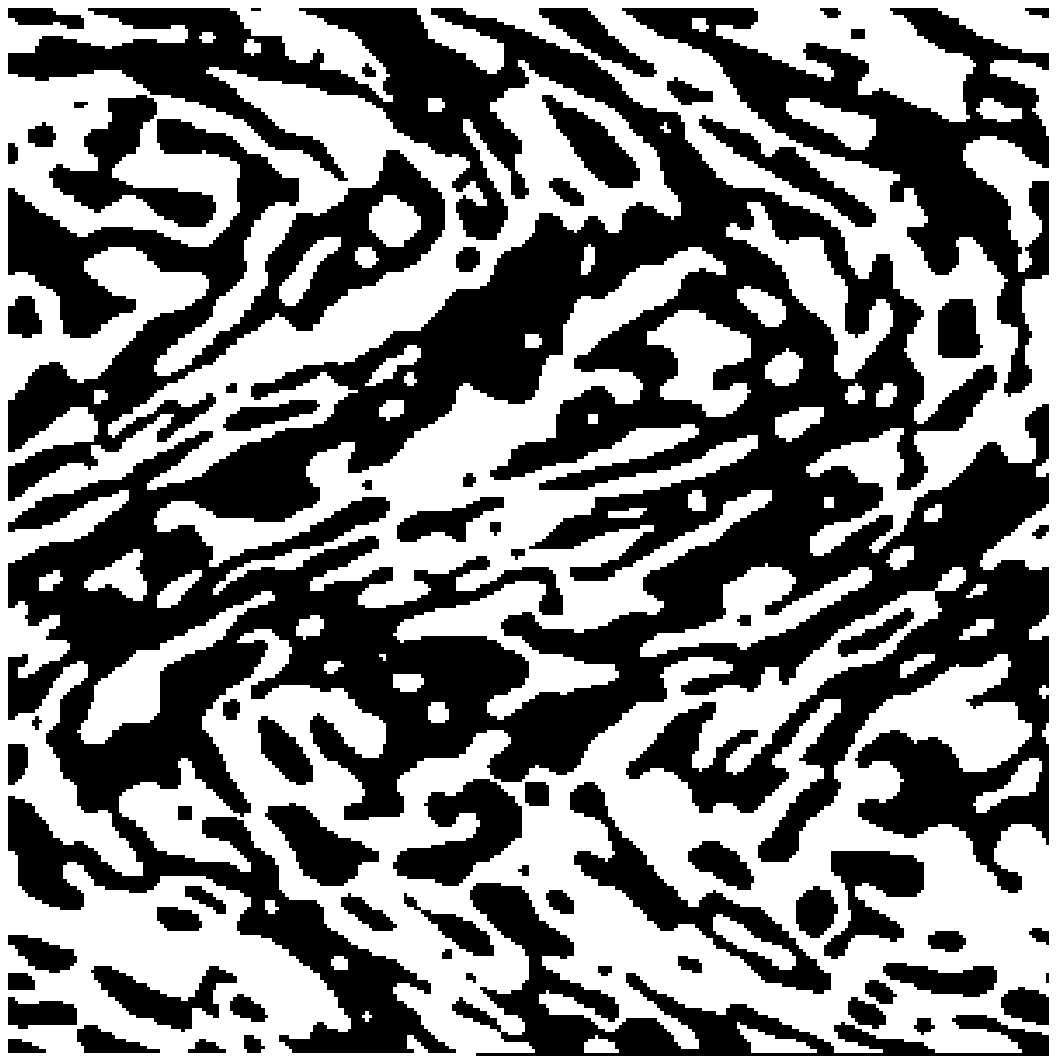,width=3.5cm,height=3.3cm} &
\psfig{file=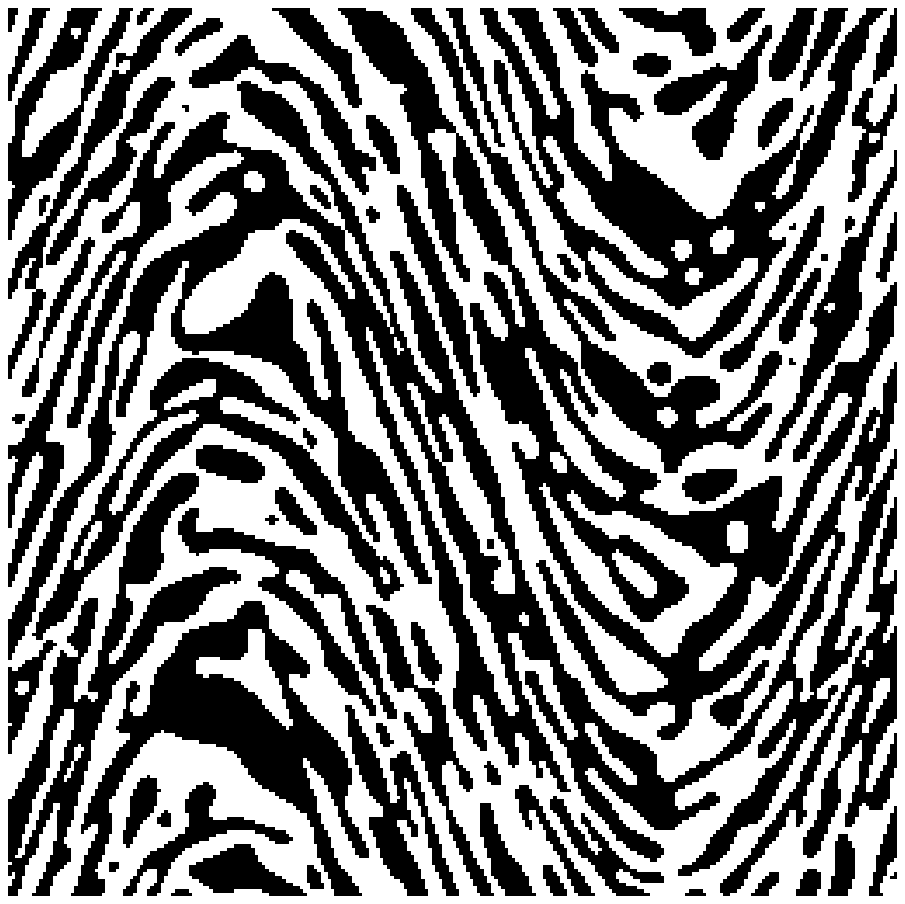,width=3.3cm,height=3.3cm}
\end{tabular}
\caption{Late stages of coarsening in the velocity
  fields of Figure \ref{field}, in the same order.}
\label{alphas}
\end{center}
\end{figure}

{\it Existence of a stationary state. ---}
We first show that a purely chaotic flow  does
indeed  stop  the domain
growth. In Fig.~\ref{secasse}, we show the evolution of
a phase-separated  sample with $\langle \phi \rangle =1/2$, 
upon turning on a 
chaotic velocity field
($\alpha=0.4$).
The  large droplets of the initial configuration are broken into
smaller droplets, until
a  stationary state where
droplets successively grow and break is reached.
Fig.~\ref{alphas} shows the late stages of coarsening of
a system with equal concentrations of phases
($\langle \phi \rangle=0$) in the four
velocity fields of Fig.~\ref{field}.
For $\alpha=0.1$, the velocity field is laminar.
We observe in that case structures very similar to those
found in a  homogeneous shear flow, but which now follow
the winding  flow lines.
In the mixed case, $\alpha=0.25$, large-domain structures
form in the laminar regions of the flow, and break into very small
domains in the chaotic ones.
In the fully
chaotic situation, $\alpha \gtrsim 0.4$, a dynamical
stationary state is reached,  with small domains
continuously breaking and reforming.
For $\alpha =1.0$, the sinusoidal nature of the underlying velocity
field becomes apparent.
This snapshot nicely illustrates the typical
`stretch and fold' processes characteristic of chaotic 
advection \cite{ottino}.

{\it Scaling properties in the chaotic flow. ---}
In the stationary regime, it is clear from Figs.~\ref{secasse} 
and \ref{alphas}
that there exist a typical length scale $L^\star$ which depends
on the parameters $D$ and $\lambda$:
this will be confirmed below by a quantitative analysis. 
As in the pure coarsening case \cite{alan1},
scaling properties are expected in the regime
$\xi \ll L^\star(D,\lambda) \ll L$.
The length scale $L^\star$ may be estimated by the following simple argument.
In the absence of flow, the domains grow as 
$L(t) / \xi \simeq  ( D t/T)^{1/3}$, and
this growth is stopped by the chaotic flow which introduces a
time scale $\lambda^{-1}$. 
Hence, we estimate $L^\star \simeq L(t = \lambda^{-1})$ and  predict
\begin{equation}
L^\star(D,\lambda)
\simeq \xi \left( \frac{D}{\lambda T} \right)^{1/3}.
\label{typical}
\end{equation}

\begin{figure}
\begin{center}
\psfig{file=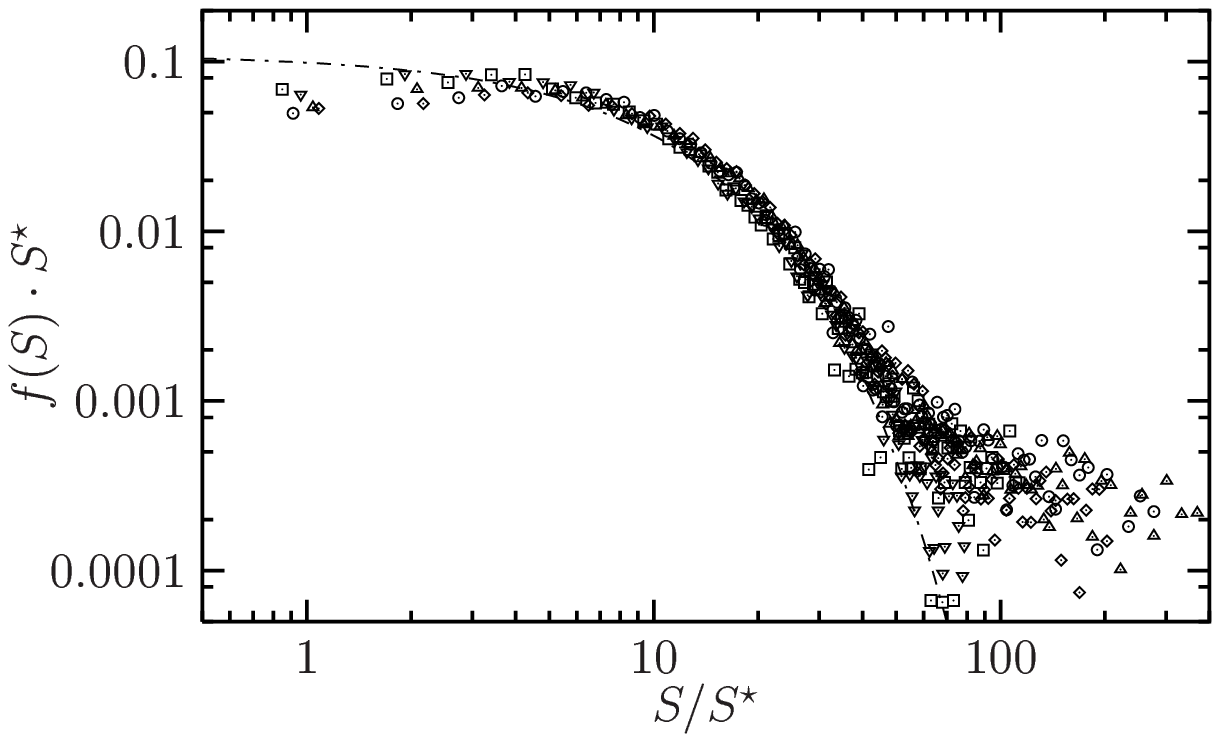,width=8.5cm,height=6cm}
\caption{Surface distribution of the droplets rescaled according
to Eq.~(\ref{scale}), where the typical surface $S^\star$ is given
by $S^\star \simeq (D/\lambda)^{0.62}$. The dotted-dashed line
is a fit to an exponential form, and the data are for a range $\alpha \in
[0.4,3.0]$ and $D\in[100,2000]$.}
\label{gouttes}
\end{center}
\end{figure}

{\it Isolated droplets, $\langle \phi \rangle \neq 0$. ---}
Following Ref.~\cite{muzio}, we characterize the assembly of droplets
by computing the distribution of droplet surfaces $f(S)$, where
$f(S)\upd S$ is the probability that the surface occupied by a droplet
is between  $S$ and $S+\upd S$. In the scaling regime, we 
expect this distribution
to be of the form
\begin{equation}
f(S) \simeq \frac{1}{S^\star} {\cal F} \left( \frac{S}{S^\star} \right),
\label{scale}
\end{equation}
where $S^\star$ is a typical droplet area. In Fig.~\ref{gouttes},
data obtained for a wide range for the values 
of $D$ and $\lambda$ are collapsed
by using a reduced variable $S/S^\star$, with
$S^\star \simeq (D/\lambda)^{0.62}$. The data collapse
is satisfactory, and the result for $S^\star$ reasonably close to what
would be expected from Eq.~(\ref{typical}), {\it i.e.} 
$S^\star \simeq (D/\lambda)^{2/3}$.
Finding an exponent slightly smaller than the one expected theoretically
is not surprising, since the typical  domain sizes are rather
small ($S^\star \lesssim 50 \xi^2$), so that the asymptotic value for the
domain growth exponent in the absence of flow
may not be reached.
The rescaled distribution functions exhibit an exponential tail,
 ${\cal F}(y) \sim e^{-y}$ (dotted-dashed line in Fig.~\ref{gouttes}).
 Such  distributions are very  is similar to those found in the experiments
 of Ref. \cite{muzio}. For the largest droplets, deviations from
 the exponential fit are observed, indicating either insufficient statistics
 or a different scaling behaviour for the extreme values of $S$ .
\begin{figure}
\begin{center}
\psfig{file=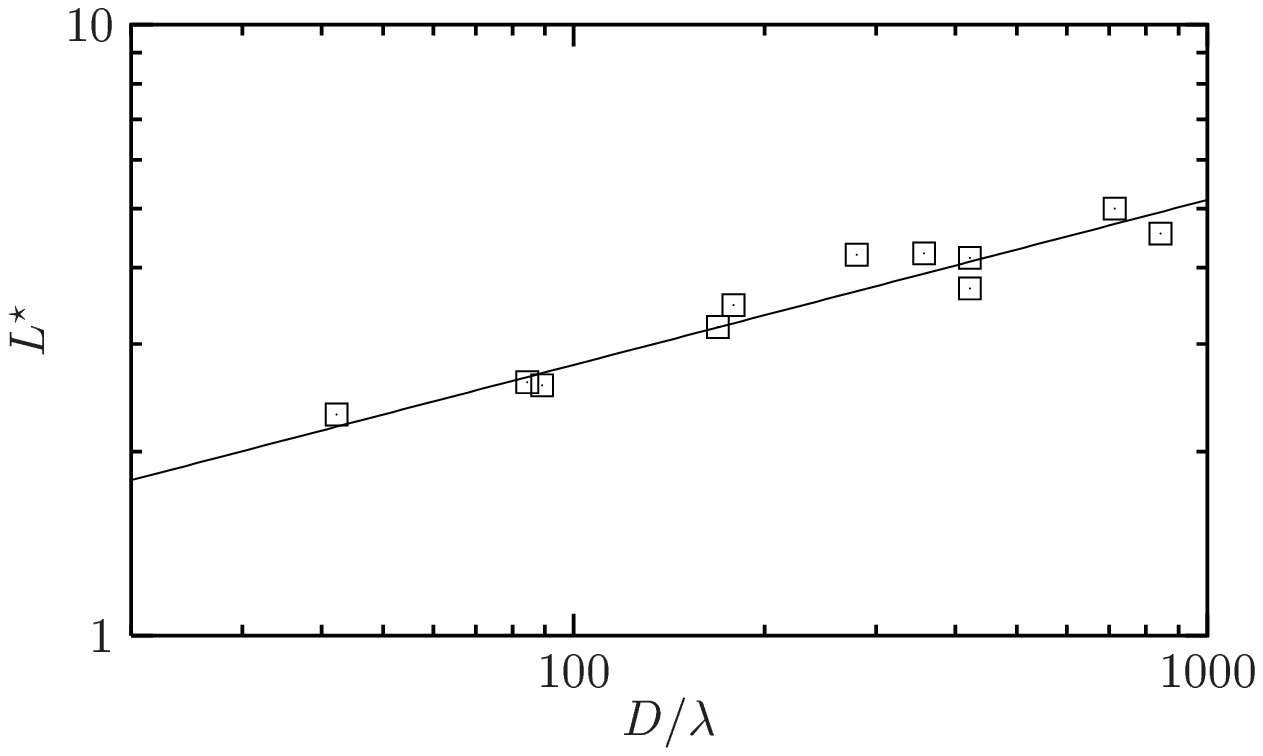,width=7.5cm,height=5cm}
\caption{Log-Log plot of the typical domain size $L^\star$ as a function
of the ratio $D/\lambda$
for the same values for $D$ and $\alpha$ as in Fig.~\ref{gouttes}.
The full line has a slope $0.27$. }
\label{lstar}
\end{center}
\end{figure}

{\it Equal concentrations: $ \langle \phi \rangle  = 0$. ---}
In the case of equal concentrations, the domains are ramified and
extend throughout the sample, so that the area is not a useful
measure of domain size. A characteristic domain size can nevertheless
be obtained from the two-point correlation function
$
C(\boldsymbol{r},t) \equiv L^{-2}
\int \upd^2 \boldsymbol{x} \langle \phi(\boldsymbol{x},t)
\phi(\boldsymbol{x+r},t) \rangle$,
which is the Fourier transform of the structure factor
measured in light scattering experiments.
Performing a time average over many configurations shows 
that the bicontinuous structure is on average perfectly isotropic,
as it is in the absence of flow.
One can therefore average $C(\boldsymbol{r},t)$ over orientations to obtain
a one variable function, $C(r)$.   
The characteristic domain size  $L^\star$
can be defined by $C(L^\star)=0.5$. 
Fig.~\ref{lstar} displays
this domain size for various combinations of $D$ and $\lambda$, as a function
of the ratio $D/\lambda$. The data can be fitted by
$L^\star
\simeq \xi \left( D /\lambda T \right)^{0.27}$ 
%\label{scalL}
Again, this is in reasonable agreement with the
scaling analysis, Eq.~(\ref{typical}). Larger simulations,
with smaller values of $\lambda$, would be necessary to
obtain larger domain sizes and avoid the crossover effects
which are well known in spinodal decomposition simulations \cite{alan1}.

\begin{figure}
\begin{center}
\psfig{file=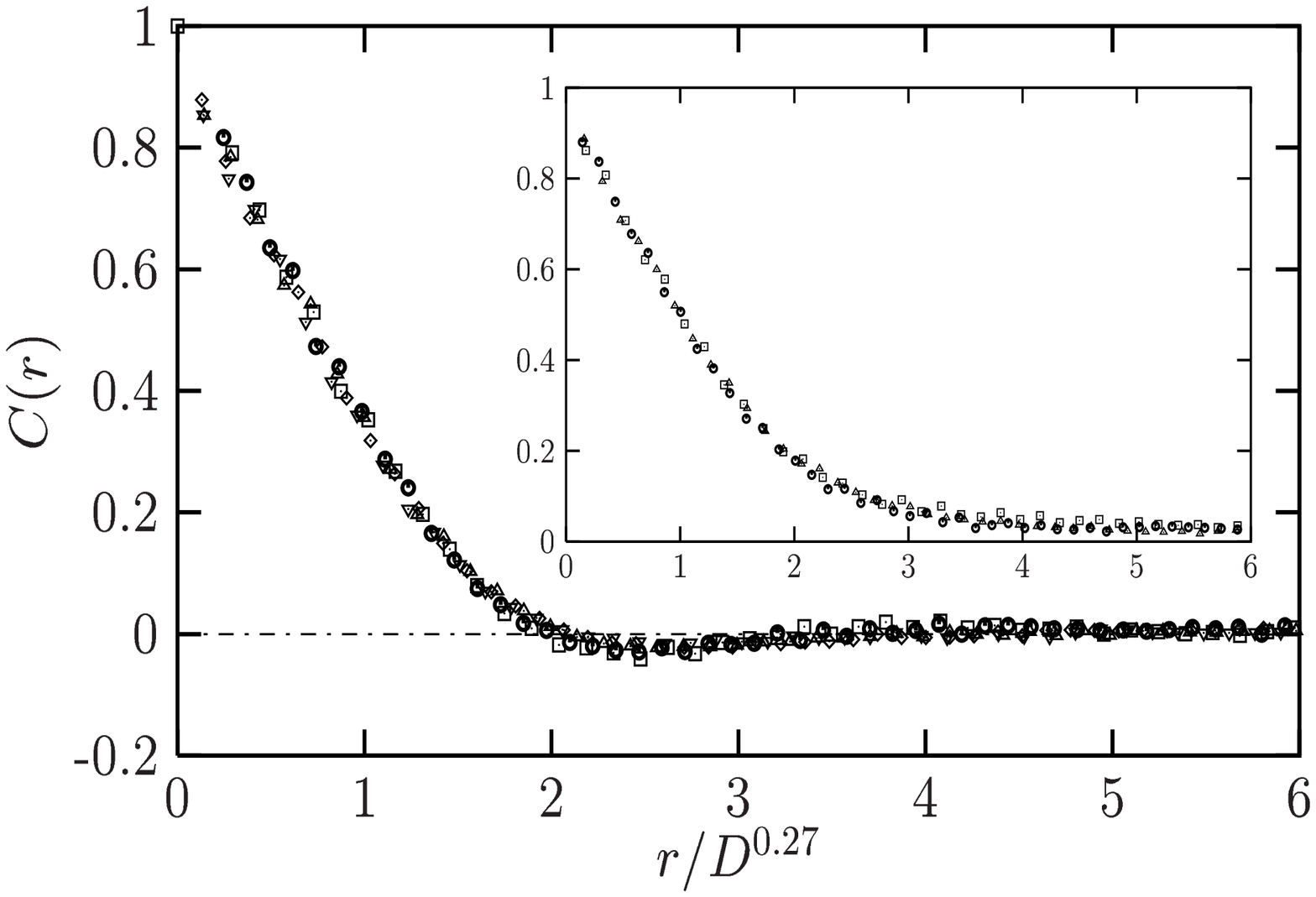,width=8.5cm,height=6cm}
\caption{Two-point correlation function $C(r)$ at fixed $\alpha$ for
for various mobility $D$, as a function of the rescaled variable
$r/D^{0.27}$. MAIN: $\alpha=1.0$ and $D=100$, 200, 400, 1000 and 2000.
INSET: $\alpha=0.4$ and $D=100$, 200 and 400.}
\label{D}
\end{center}
\end{figure}

A more detailed information on the domain structure is
obtained from the full correlation function  $C(r)$. Here, one could expect
from the scaling hypothesis a behaviour of the form
\begin{equation}
C(r) \simeq {\cal C} \left(\frac{r}{L^\star}\right)
\label{scal}
\end{equation}
with ${\cal C}$ a universal function. This hypothesis is
tested in Fig.~\ref{D}, where $C(r)$ is represented
for fixed $\lambda$ and various
values of $D$.   At fixed $\lambda$, a good collapse of the data
obtained for different $D$ is achieved by using a rescaled variable
$r/D^{0.27} $.
The inset of Fig.~\ref{D}, however, shows that the shape
of the scaling function slightly depends on $\lambda$, 
so that the universal
scaling expressed by Eq.~(\ref{scal}) is not valid. 
We attribute this change of the scaling
function with the flow pattern to the fact that 
even in the chaotic regime the flow cannot be considered as
being homogeneous and isotropic,
but exhibits an underlying sinusoidal structure.
This is in contrast with the case of isolated droplets where
the droplet distribution was not affected by this structure, recall
Fig.~\ref{gouttes}. 

We have studied the phase separation in conditions in which the 
species boundaries are passively advected by a incompressible flow.
We have shown that a chaotic flow results in a steady
state with domains of finite size
resulting from the balance
between spinodal decomposition and
 chaotic advection, Eq.~(\ref{typical}). This should be contrasted with the
situation observed in turbulent flow,
 where the flow intensity must exceed a threshold 
in order to stop domain growth
\cite{esp}. Such a difference can be traced back to the fact that Lyapounov
exponents for passive scalar advection are actually 0 in the latter case.
 The essential approximation  in our work, compared to
realistic experimental situations,  is the assumption that the flow pattern is
not modified by the  domain growth. This assumption, however, may not be
unrealistic  if the two fluids have similar viscosities and if the capillary 
stresses are small compared to viscous stresses. This is measured by  the
capillary number $C_a= \eta \lambda / (\gamma/L^\star)$, where  $\eta$ is the
viscosity and $\gamma$ the surface tension.  In highly viscous  fluids, $C_a$
is expected to be large, so that the decoupling is possible.  This decoupling
also makes it possible to consider analytical treatments.

We  acknowledge useful discussions with A. J. Bray,  B.
Cabane, P. Leboeuf, J. F. Pinton and J. E. Wesfreid.


\begin{references}
\bibitem{foot}Droplet formation in three dimensions  at high dilution 
may happen at times long enough
to destabilize cilindrical domains, but too short to form lamellar domains.
\end{references}
\end{document}